\def\BEQ{\begin{eqnarray}}
\def\EEQ{\end{eqnarray}}
\def\BML{\begin{mathletters}}
\def\EML{\end{mathletters}}
\def\BE{\begin{equation}}
\def\EE{\end{equation}}
\def\NN{\nonumber}
\def\G{\Gamma}
\def\jv{{\bf j}}
\def\phib{\skew5\bar\phi}
\def\a{\alpha}
\def\b{\beta}
\def\e{{\rm e}}
\def\s{\sigma}
\def\sb{\mbox{\boldmath$\sigma$}}
\def\xib{\bar\xi}
\def\eps{\varepsilon}
\def\d{\partial}
\def\dg{\dagger}
\def\up{\uparrow}
\def\dn{\downarrow}
\def\hf{{1\over2}}
\def\Tr{{\rm Tr\;}}
\def\rep#1{{\bf #1}}

\documentstyle[aps,prb,floats]{revtex}
\input epsf
\begin{document}

\title{SO(5)-Symmetric Description of the Low
Energy Sector of a Ladder System}
\author{David G. Shelton\cite{Shelton} and David S\'en\'echal}
\address{Centre de Recherche en Physique du Solide et D\'epartement de Physique,}
\address{Universit\'e de Sherbrooke, Sherbrooke, Qu\'ebec, Canada J1K 2R1.}
\address{{\tt CRPS-97-24  |  cond-mat/9710251}~~\rm(revised version, March
1998)}
\maketitle

\begin{abstract}
We study a system of two Tomonaga-Luttinger models coupled by a small
transverse hopping (a two-chain ladder). We use Abelian and non-Abelian
bosonisation to show that the strong coupling regime at low energies can be
described by an SO(5)$_1$ WZW model (or equivalently 5 massless Majorana
fermions) deformed by symmetry breaking terms that nonetheless leave the theory
critical at $T=0$. The SO(5) currents of the theory comprise the charge and
spin currents and linear combinations of the so-called pi operators (S.C.
Zhang, Science 275, 1089 (1997)) which are local in terms both of the original
fermions and those of the effective theory. Using bosonisation we obtain
the asymptotic behaviour of all correlation functions. We find that the 5
component ``superspin'' vector has power law correlations at $T=0$; other
fermion bilinears have exponentially decaying correlations and the
corresponding tendencies are suppressed. Conformal field theory also allows us
to obtain the energies, quantum numbers, and degeneracies of the low-lying
states and fit them into deformed SO(5) multiplets.
\end{abstract}
%
\widetext

\section{Introduction}

One of the most characteristic features of the high-T$_c$ Cuprates is the
proximity of antiferromagnetic (AF) and superconducting (SC) phases as a
function of doping. As a result, much of the theoretical effort has
focused on trying to consistently treat the
insulating-underdoped-optimally doped region of the phase diagram, in
which AF and SC tendencies compete and may have strong fluctuations.

An interesting recent proposal is that of Zhang.\cite{zhang} He
suggests that the simplest way of unifying AF and SC in the Cuprates is
to introduce a new five-component vector order parameter consisting of
the three component staggered magnetisation, and two components 
associated with the real and imaginary parts of the d-wave SC order
parameter. Clearly this new concept is only useful if there exists some
kind of symmetry (higher than the known SO(3)$\otimes$U(1)) which
relates the AF and SC sectors. His suggestion is that an approximate
SO(5) symmetry emerges in the low energy sector (SO(5) because the new
composite order parameter has five components and transforms like a
vector). If true, this would allow the construction of an SO(5) quantum
nonlinear $\sigma$-model to explain the low-energy dynamics of the high
T$_c$ materials. This could explain the form of the phase diagram, and
the so-called $\pi$-mode.\cite{demler}

However, there have been several criticisms of this theory. Some 
\cite{greiter} have focused more on the details of microscopic
calculations in the framework of the t-J or Hubbard models. Others have
added several physical objections.\cite{baskan} One response to these
criticisms has been to attempt to construct concrete examples of
extended microscopic Hamiltonians which manifestly have an SO(5)
symmetry.\cite{microscopic} But knowing the Hamiltonian does not
necessarily tell us much about the low energy behaviour.

In this paper we study a two-chain ladder Hamiltonian that is related 
to popular two-dimensional models of the Cuprates. One of the reasons
that ladder systems have attracted such attention is that many
experimental realisations of these systems are very closely related to
the high T$_c$ materials \cite{lad}, and some have even exhibited
superconductivity.\cite{ueh,iso} From a theoretical
point of view, powerful non-perturbative techniques such as bosonisation
and conformal field theory (CFT) exist in one dimension. This offers
hope of starting with a microscopic  Hamiltonian and ending up with a
tractable effective field theory. In this paper it is not our purpose to
comment on the general validity of the SO(5) idea but to explicitly
study a simplified and more tractable model.

There is a large body of literature on two-chain and ladder systems
\cite{dag}$^-$\cite{fabrizio} (for a review see Ref.~\onlinecite{lad}).
 Using a combination of weak coupling RG and bosonisation, the phase
diagram has been intensively investigated. These analyses reveal that
for small interchain hopping there are interesting strong coupling
phases. However, whilst Abelian bosonisation and weak coupling RG are
good for determining the phase diagram, they do not explicitly respect
the symmetries of the system, nor do they provide detailed information
about the correlations.
In this paper we explore in more detail the strong coupling region of
a two chain ladder system, taking care to preserve the full non-Abelian
symmetries and obtain the correlations.

It is well known that many 2 chain ladder systems are spin liquids;
that is, they exhibit a spin gap for a wide range of different
fillings and couplings. This is because the Luttinger liquid is a
quantum critical system, and as such, highly unstable to perturbations
such as interchain coupling. In general there are a number of relevant
couplings which can drive the system into a spin gap phase (an
explicit example is discussed in section VI). However, in this paper
we study a simplified system in which there is no backscattering and
as a result, no spin gap. This model is of interest because it
displays remarkable similarities to some aspects of the Zhang proposal
in 2 dimensions.\cite{zhang}

The model we consider is a system of 2 spinful Tomonaga-Luttinger (TL)
models in the repulsive regime, coupled by a small interchain hopping.
This corresponds to the case of no backscattering and was studied in
Refs.~\onlinecite{fink},\onlinecite{schulz},\onlinecite{fabrizio}.
We demonstrate that the hopping only generates couplings in a certain
sector of the theory (which we call ``flavour''), freezing it out of the
effective action at energy scales below $t_{\perp}$. In agreement with
the above references, we find that this leaves a critical (at T=0) spin
and charge sector with conformal charge 5/2. However, we go on to show
that this can be represented as a system of 5 massless Majorana
fermions, or equivalently, an SO(5)$_1$ Wess-Zumino-Witten (WZW) model,
deformed away from the symmetric point by marginal current-current
interactions. These SO(5) breaking terms are associated with spin charge
separation (spin and charge velocities not equal $v_s\neq v_c$) and the
anomalous charge exponent ($K_c\neq 1$), which distinguish the spin and
charge sectors.  Thus the system is never exactly SO(5) symmetric except
in the trivial noninteracting case. Nonetheless, this representation
does have strong analogies with the Zhang proposal in 2d; the physics
can be understood using an SO(5) symmetric $\sigma$-model with symmetry
breaking terms. In this way we obtain the asymptotic behaviour of all
correlation functions; the correlations of the 5 component ``superspin''
are enhanced (power law at T=0); we obtain their scaling dimensions.
Other fermion bilinears die away exponentially fast.

Sections II-V are concerned with an analysis of this model, including
its detailed symmetric description, the relevant currents, the
$\pi$-operators, its correlations and low lying multiplets in the
excitation spectrum. One important way in which the system we are
studying differs from that considered by Zhang is that we are away
from half-filling, which is a very special point in 1d.
Exactly at half-filling it is necessary to consider the
Umklapp term, which causes a Mott gap in the charge sector.
\cite{giamarchi} Then the low energy effective Hamiltonian is simply a
pure spin Heisenberg model (with exchange
$J\sim 4t^2/U$ in the case of the repulsive Hubbard model at strong $U$). We comment
further on this difference in section IV.

In section VI we finally consider the case of two coupled Luttinger
liquids, which differs from the previous model in that it includes
marginal backscattering terms. An example of this is provided by some
regions of the phase diagram of a system of 2 Hubbard chains coupled
by single particle hopping. In this more physical case, we show in
detail how the additional marginal terms cause a spin gap to appear in
agreement with Refs.~\onlinecite{dag}-\onlinecite{fabrizio},
and numerical work such as Ref.~\onlinecite{scalapino}.
Then the spectrum and correlations are as in Ref.~\onlinecite{dgs};
there is a spin gap but the charge sector remains gapless.

Finally, we conclude. There is also an appendix which sketches out a
bosonisation prescription that enables us to calculate the correlation
functions of fermion bilinears.

\section{A Simple Model}

Many systems of interacting one-dimensional fermions away from half filling
fall into the Luttinger liquid universality class. That is, they exhibit
spin-charge separation, gapless excitations, anomalous power law correlations
and the absence of a quasiparticle pole (see Ref.~\onlinecite{voit} for a
recent review, and references therein). For example, the one-dimensional
repulsive Hubbard model away from half-filling is known from its exact solution
to be a Luttinger liquid all the way from $U=0$ to $U=\infty$, as is the t-J
model for small enough $J/t$.

One of the simplest two-chain models of this type that can be written down 
consists of two Tomonaga-Luttinger (TL) models (labelled by a chain index
$i=1,2$) coupled by a small transverse hopping $t_{\perp}\ll t$:
\BE
\label{TLmodel}
H =  H_{\rm TL}(1)+H_{\rm TL}(2)+H_\bot
\EE
where the TL Hamiltonian is a sum of three pieces ($H_0+H_2+H_4$):
\BEQ
H_0(i)=& iv_F&\sum_\a\int dx\;  
\left(R^\dg_{\a,i}\d_x R_{\a,i}-L^\dg_{\a,i}\d_x L_{\a,i}\right) \NN\\
H_2(i)=& g_2&\sum_{\a,b'}\int dx\;
j_{\a,i}^R(x)j_{\b,i}^L(x)\NN\\
H_4(i)=& g_4&\sum_{\a,\b}\int dx\;
\left( j_{\a,i}^R(x)j_{\b,i}^R(x)
+ j_{\a,i}^L(x)j_{\b,i}^L(x)\right)
\label{tl}
\EEQ
The current (or density) is simply defined as
\BE\label{density}
j^R_{\a,i} = R_{\a,i}^\dg R_{\a,i} \qquad
j^L_{\a,i} = L_{\a,i}^\dg L_{\a,i} 
\EE
and the electrons fields $R_{\a,i}$ and $L_{\a,i}$ are slowly varying on an
atomic scale: the electron annihilation operator at site $x$, chain $i$ and
spin $\a$ may be expressed as
\BE\label{cont}
c_{\a,i}(x) =  R_{\a,i}(x)\e^{ik_Fx} + L_{\a,i}(x)\e^{-ik_Fx}
\EE
In terms of these fields, the simple interchain hopping term becomes
\BE
H_\bot = t_{\perp}\int dx\; \sum_\a\left( 
R_{\a,1}^\dg(x)R_{\a,2}(x) + L_{\a,1}^\dg(x)L_{\a,2}(x) + {\rm h.c.}
\right)
\EE
For simplicity, we have assumed that the Hamiltonian is invariant under spin
rotation, and so the coupling constants $g_2$ and $g_4$ are the same for
parallel and antiparallel spin configurations. Normal ordering is assumed
throughout in products of local fields (definition of currents, Hamiltonians,
etc.).

It is worth making a quick observation about the difference between the terms
TL {\it liquid} and TL {\it model}:  The TL model is an idealised and specific
Hamiltonian, written down in  Eq.~(\ref{tl}). It has a perfectly linear
dispersion, an infinitely deep Fermi sea, has only density-density interactions
and is exactly solvable for all values of the coupling constants (the model is
unstable beyond a critical value of $g_2$).\cite{voit} The TL liquid (which is
the generic state corresponding to many realistic Hamiltonians like the Hubbard
model away from half filling) differs in that the dispersion is no longer
exactly linear, and the Fermi sea no longer infinitely deep. But from our point
of view the most important difference in the low energy sector is the presence
of marginally irrelevant couplings (backscattering).
 In a single chain system these are not very
important when repulsive -- they simply give logarithmic corrections to the
correlation functions. In section VI we will study the effect of these
additional terms in the two chain system, in order to establish the behaviour
of the more realistic coupled TL {\it liquids}, but for the moment, we will
restrict our attention to the simpler case of coupled TL {\it models}. 

The model (\ref{TLmodel}), even though it is made of TL models, is not exactly
solvable because of the interchain hopping. However, we will argue presently
that the model segregates into three different {\it sectors}, respectively
associated with charge, spin, and ``flavour'', and that the combined
effect of interchain hopping and interactions is to make the flavour
sector massive, leaving only the charge and spin sectors critical (i.e.,
gapless). To each sector one may associate current operators, expressed as
bilinears of the electron fields:
\BEQ
\label{currents}
\hbox{charge:}\qquad &J_R(x) &= \sum_{\a,i} R^\dg_{\a,i}(x)R_{\a,i}(x)\NN\\
\hbox{spin:}\qquad &{\bf J}_R(x) &= \hf\sum_{i,\a,\b} R^\dg_{\a,i}(x)
\sb_{\a\b} R_{\b,i}(x)\NN\\
\hbox{flavour:}\qquad &{\bf I}_R(x) &= \hf\sum_{i,j,\a} R^\dg_{\a,i}(x)
\sb_{ij} R_{\a,j}(x)
\EEQ
where $\sb$ is the vector of Pauli matrices (left-moving
currents are defined similarly). These currents have the following commutation
relations (they may be derived from Wick's theorem):
\BEQ
[J_R(x),J_R(y)] &=& -{2i\over\pi}\delta'(x-y) \cr
[J_R^a(x),J_R^b(y)] &=& -{i\over 2\pi} \delta^{ab}\delta'(x-y)
+ i\eps^{abc}J_R^c(y)\delta(x-y) \cr
[I_R^i(x),I_R^j(y)] &=& -{i\over 2\pi} \delta^{ij}\delta'(x-y)
+ i\eps_{ijk}I_R^k(y)\delta(x-y) 
\EEQ
and currents of different types (i.e., charge, spin, and flavour)
commute.
Thus in the language of non-Abelian bosonisation, the charge current
obeys a U(1) Kac-Moody algebra, and the spin and flavour currents obey
SU(2)$_2\equiv$SO(3)$_1$ algebras.\cite{book,zamfat}
It is simple to show that the Hamiltonian (\ref{TLmodel}) may be
expressed as $H=H_0+V_c+V_f$, where only the above currents
appear. This is just a matter of taking careful account of
point-splitting and normal ordering:\cite{stone}
\BEQ
\label{ham2}
H_0 &=& {\pi v_F\over 2}\int dx\left(
J_R^2 + {\bf J}_R^2 + {\bf I}_R^2 + ~[R\to L] \right) \NN\\
V_c &=& \hf\int dx\left\{ g_2 J_RJ_L+g_4(J_R^2+J_L^2)\right\} \NN\\
V_f &=& 2\int dx\left\{ g_2I_R^zI_L^z+g_4[(I_R^z)^2+(I_L^z)^2]
+ t_\bot ( I_R^x + I_L^x ) \right\}
\EEQ
Therefore the model (\ref{TLmodel}) decouples into three independent sectors
(charge, spin, flavour). The important point is that the hopping
term only involves the flavour sector, which is decoupled from
the other two. The effect of interactions ($g_2$ and $g_4$) on
the charge sector will be a velocity renormalisation and anomalous
scaling exponents ($K_c\ne1$). The combined effect of interactions and
transverse hopping on the flavour sector is more dramatic. The RG
analysis of Ref.~\onlinecite{fabrizio} shows unambiguously that in the
repulsive regime ($K_c<1$), the system scales to strong coupling at
energies $<t_{\perp}$ (in the notation of Ref.~
\onlinecite{fabrizio} our model
corresponds to initial conditions of $g_i^{(1)}=0$, $g_i^{(2)}=-g_i^{||}=g_0$
for $i=0,\pi,f,t,b$). The combination of the small 
hopping term $t_{\perp}$ and the
interaction terms leads to the generation of important couplings in the RG
process, giving a gap in some channels. What our analysis tells us is that all
of this physics is only happening in the flavour sector, and thus it is this
sector that becomes  gapped, while the total spin and total charge sectors
remain untouched and critical. So at low enough energies the flavour sector is
frozen out of the effective theory, and our task is simply to understand the
remaining charge and spin degrees of freedom.

\section{Spinor and vector descriptions}

Each electron field $R_{\a,i}$ or $L_{\a,i}$ carries charge, spin and
flavour. The separation of the model into charge, spin and flavour
sectors is therefore difficult to describe in terms of these operators.
However, one may introduce a different set of Fermi fields in terms of
which this separation is much more natural. To this end, we must use
some representation theory of Lie groups.

Let us first consider the model (\ref{TLmodel}), but without interactions or
interchain hopping (i.e., two free, decoupled chains). This model has SO(8)
symmetry, and this may be shown has follows.
Each complex field $R,L$ may be written in terms of its real and imaginary
parts: $R_{\a,i}=R_{1,\a,i}+iR_{2,\a,i}$ and then, except for a total
derivative, the Hamiltonian $H_0$ takes the form
\BE
H_0 = iv_F\sum_{\mu}\int dx\; (R_\mu \d_x R_\mu-L_\mu \d_x L_\mu)
\EE
where the composite index $\mu$, running from 1 to 8, stands for spin,
chain and real/imaginary part. The eight Fermi fields $R_\mu$ (or
$L_\mu$) can undergo an internal SO(8) rotation that leaves
$H_0$ invariant. Hence the model has a chiral SO(8) symmetry. It is well
known that a collection of $N$ real free fermions like this is equivalent to a
special kind of conformal field theory: a level-1 SO(N) WZW
model.\cite{DMS} Chiral
SO(8) currents may be defined in terms of those real fermions as follows:
\BE\label{so8currA}
J^A_R = \hf\sum_{\mu,\nu=1}^8 R_\mu S^A_{\mu\nu} R_\nu
\EE
where $S^A_{\mu\nu}$ is a matrix representation of the generators of SO(8) ($A$
runs from 1 to $\hf N(N-1)=28$, the number of generators). Left-moving currents
are defined similarly. The charge, spin and flavour currents (\ref{currents})
are special cases of the above and correspond to specific values of the index
$A$ if the generators $S^A_{ij}$ are chosen judiciously.

\begin{figure}
\epsfxsize=6cm\centerline{\epsfbox{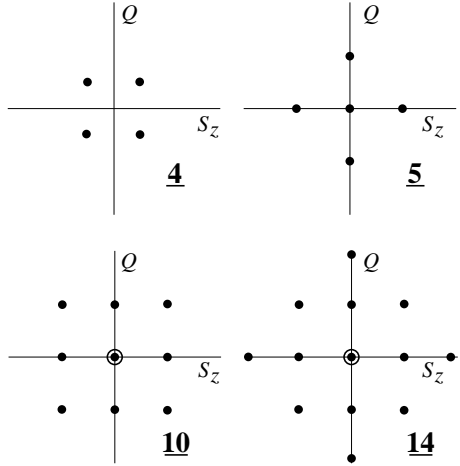}}\vglue4mm
\caption{The $S_z-Q$ diagrams associated with the lowest nontrivial
SO(5) multiplets.}
\label{so5fig}
\end{figure}

The currents (\ref{so8currA}) are bilinears in the electrons fields $R_\mu$
($\a=1,\ldots,8$). However, the SO(8)$_1$ WZW model contains other fields,
belonging to a different representation of SO(8), in terms of which these
currents are also bilinears. Among all SO(N) groups, SO(8) is peculiar in that
its vector representation, of dimension 8, has properties identical to its
spinor and conjugate spinor representations (also of dimension 8). Indeed,
which one is called `vector' is a matter of convention, dictated by the way the
SO(8) symmetry breaks down to smaller SO(N) components. In order to decide to
which SO(8) representation the electron fields belong, one must study in detail
how each representation breaks down when the symmetry is reduced. Let us
consider a two-stage symmetry breaking, in which the flavour sector, with its
SU(2), is first segregated, and then the charge U(1) and spin SU(2) (note that
U(1)$\sim$SO(2) and SU(2)$\sim$SO(3)):
\BE\label{break}
SO(8) \to SO(5)\otimes SO(3)^{\rm fl.} \to
SO(2)^{\rm c}\otimes SO(3)^{\rm sp.}\otimes SO(3)^{\rm fl.}
\EE
We stress that the goal of the present analysis is to fit the fields and states
of the model into symmetry multiplets, without demanding the symmetry to be
exact. In the first stage of this breakdown, the vector and spinor
representations of SO(8) are decomposed as follows (irreducible representations
will be commonly denoted by bold numbers giving their dimensions, with an
occasional superscript distinguishing between vectors ($v$) and spinors ($s$)):
\BEQ
\rep8^v &\to& (\rep5,\rep1) ~\oplus~(\rep1,\rep3) \NN\\
\rep8^s &\to& (\rep4,\rep2)
\label{decomp1}
\EEQ
(here, for instance, the notation $(\rep4,\rep2)$ stands for a tensor product of
the 4-dimensional representation of SO(5) with a doublet of SU(2)$^{\rm fl.}$).
Since SO(5) representations are not all that familiar, we provide a pictorial
view of the lowest nontrivial ones on Fig.~\ref{so5fig}. The multiplet $\rep4$
is the spinor representation of SO(5), while $\rep5$ is the vector
representation and $\rep{10}$ the adjoint representation, i.e., the
representation of the SO(5) symmetry currents or generators. The decomposition
of these SO(5) representations in terms of spin multiplets and charge quantum
numbers is best appreciated on Fig.~\ref{so5fig}. For instance, the SO(5)
spinor $\rep4$ breaks down into two spin-$\hf$ doublets, one with charge +1 and
the other with charge -1. On the other hand, the vector representation breaks
down into a spin-1 triplet of charge zero and two singlets of charges $\pm2$.

We may now ascertain that the electron fields $R_\mu$ belong to the
{\it spinor} representation of SO(8). Indeed, the lowest excited states of
$H_0$, obtained by acting on the vacuum with the lowest electron creation
operators, form a multiplet of 4 states of charge $+1$ and four states of
charge $-1$. This is precisely the charge content of the spinor multiplet
$\rep8^s$, since the spinor $\rep4$ of SO(5) contains two states of charge $+1$
and two of charge $-1$, and appears twice in the decomposition (\ref{decomp1}),
because of the flavour doublet.

A different set of real fermions, denoted $\xi_i$ ($i=1,\ldots,8$), belongs to
the {\it vector} representation of SO(8).
These new fermions are related in a complicated, nonlocal way to the
original fermions. The transformation relating them may be explicitly
obtained via Abelian bosonisation, if one takes care to preserve the
anticommutation factors, but this is not a particularly
illuminating procedure. The important point is that they are just a different
basis or representation for the same system.
 These fermions obey the usual
anticommutation relations $\{\xi_i(x),\xi_j(y)\}=\delta_{ij}
\delta(x-y)$. The
SO(8) currents (\ref{so8currA}) may also be expressed as bilinears of these
fermions, albeit with the help of a different set of SO(8) matrices:
\BE
J_R^A = {1\over 2}\sum_{ij}\xi_iT^A_{ij}\xi_j 
\EE
A characteristic
 feature of the vector representation is its particularly simple
decomposition into charge, spin and flavour components: in the first stage of
the breakdown (\ref{break}), the vector representation decomposes as
$\rep8^v\to (\rep5,\rep1) \oplus (\rep1,\rep3)$. In the second stage,
the SO(5) vector decomposes as
$\rep5\to(\rep3,\rep1)\oplus(\rep1,\rep2)$ (this time, doublets on the
r.h.s. correspond to spin and charge multiplets, respectively). We may thus
distinguish three Majorana fermions ($\xi^s_i$, $i=1,2,3$) for spin, three
others ($\xi^f_i$, $i=1,2,3$) for flavour and the remaining two ($\xi^c_i$,
$i=1,2$) for charge. The spin, flavour and charge current then have the
following expressions:
\BEQ
J_R^{i}&=&-{i\over 2}\epsilon_{ijk}\xi^s_j\xi^s_k \NN\\
I_R^i&=&-{i\over 2}\epsilon_{ijk}\xi^f_j\xi^f_k \\
J_R&=& -i\epsilon^{jk}\xi^c_j\xi^c_k = -2i\xi^c_1\xi^c_2
\label{currents2}
\EEQ
It is interesting to note that these currents (and eventual SO(5) currents) are
local in terms both of the electron fields {\it and} in terms of the above
Majorana fermions, even though the fermion operators themselves are nonlocally
related. The Majorana fermions are nonetheless legitimate operators of the
theory. For instance, at half-filling, if all sectors became gapped, the three
spin fermions $\xi^s_i$ would describe the triplet of spin excitations
characteristic of a gapped spin-1 chain.\cite{tsvelik}

\section{The SO(5) currents and Zhang's $\Pi$ operators}

We have seen above that the spin and charge degrees of freedom, which make up
the critical sector of the theory (\ref{TLmodel}), may be described by the five
Majorana fermions $\xi^c_{1,2}$ and $\xi^s_{1,2,3}$. Except for the interaction
$V_c$ of Eq.~(\ref{ham2}), the low-energy sector is equivalent to a level-1 SO(5) WZW
model with conformal charge $c={5\over2}$.\cite{schulz} Indeed, the above
Majorana fermions may be arranged in the following suggestive sequence:
\BE\label{so5vector}
\xi_1 = \xi^c_2 \qquad
\xi_2 = \xi^s_1 \qquad
\xi_3 = \xi^s_2 \qquad
\xi_4 = \xi^s_3 \qquad
\xi_5 = \xi^c_1 
\EE
plus corresponding left-moving fields. That the non-interacting part of the
spin-charge sector is equivalent to a level-1 SO(5) WZW model means that this
part of the Hamiltonian may be simply expressed as\cite{witten}
\BE
H_0 = iv_F\sum_{j=1}^5 \int dx\; (\xi_j\d_x\xi_j-\xib_j\d_x\xib_j)
\EE
(here $\xib_j$ denote the left-moving fields).

It is then useful and instructive to introduce the full ten SO(5) currents. Four
of those currents are provided by the charge and spin currents of
Eq.~(\ref{currents}). The remaining six, corresponding to Zhang's 
$\pi$-operators,\cite{zhang} may be expressed in the continuum limit either in
terms of the Majorana fermions $\xi^c_{1,2}$ and $\xi^s_{1,2,3}$, or directely
in terms of the electron fields.

It is interesting at this point to go back to the lattice definition of the
$\pi$ operators:\cite{zhang}
\BEQ\label{pidef}
\Pi^\dg_{a}&=&
\sum_{{\bf k},\a,\b} g({\bf k}) c_\a^\dg( {\bf k}+{\bf Q})
\left(\s_{a}\s_2\right)_{\a\b} c_\b^\dg(-{\bf k})\NN\\ 
&=&\sum_{{\bf m},{\bf n},\a,\b}g_{{\bf m},{\bf n}}
\e^{i{\bf Q}\cdot{\bf m}}c_\a^\dg({\bf m})
\left(\s_a\s_2\right)_{\a\b}c^\dg_\b ({\bf n})
\EEQ
where {\bf m} and {\bf n} are vectorial site indices (in-chain and chain
index). On a square lattice at half-filling
Zhang takes ${\bf Q}=(\pi,\pi)$. On a two-chain
system, away from half filling, there are two possibilities: ${\bf
Q}=(2k_F,\pi)$ for right movers and ${\bf Q}=(-2k_F,\pi)$ for left movers.

The structure factor $g({\bf k})=\cos k_x-\cos k_y$ has the local form:
\BE
g_{{\bf m},{\bf n}} = \left\{\begin{array}{rl} +2
&\mbox{if ({\bf m},{\bf n}) are NN on the same chain}\\
-2 &\mbox{if ({\bf m},{\bf n}) are NN on opposite
chains}\\ 
0&\mbox{otherwise}\\
\end{array}\right.
\label{gmn}
\EE
Defining a ``staggered $\pi$-density'':
\BE
\Pi^\dg_{a} = 8\int dx\;
{\rm e}^{2ik_Fx}\pi_a^\dg\label{stagpi}
\EE
we find, with the help of Eq.~(\ref{cont}), the following expressions:
\BEQ\label{picont}
\pi^\dg_x&=&-{i\over 2}\left[R^\dg_{\up,1}
R^\dg_{\up,2}-R^\dg_{\dn,1}
R^\dg_{\dn,2}\right]\NN\\
\pi^\dg_y&=&-{1\over 2}\left[R^\dg_{\up,1}
R^\dg_{\up,2}+R^\dg_{\dn,1}
R^\dg_{\dn,2}\right]\NN\\
\pi^\dg_z&=& \phantom{-}{i\over
2}\left[R^\dg_{\up,1}R^\dg_{\dn,2}
+R^\dg_{\dn,1}R^\dg_{\up,2}\right]
\EEQ
Interestingly, this continuum expression for the $\pi$ operators is quite robust
and does not depend too closely on the microscopic definition (\ref{pidef}).
One might have alternatively chosen the Henley-Kohno form
$g({\bf k})=$sgn$(\cos k_x-\cos k_y)$ and this would not have changed the
results, apart from derivative terms which are irrelevant in the RG sense --
essentially because in a two-chain ladder system there are only two
$k_y$ values, $0$ or $\pi$. Another seemingly different microscopic expression
for the $\pi$ operators is used in Ref.~\onlinecite{SZH}, but again, we have
verified that the same expression (\ref{picont}) is obtained in the continuum
limit.

We define the matrix $l^{ab}(x)$, analogously to Zhang:
\BE
\left(\begin{array}{ccccc} 0&&&&\\
\pi^\dg_x+\pi_x&0&&&\\
\pi^\dg_y+\pi_y&-J^z_R&0&&\\
\pi^\dg_z+\pi_z&J^y_R&-J_R^x&0&\\
J_R&-i(\pi^\dg_x-\pi_x)&-i(\pi^\dg_y-\pi_y)
&-i(\pi^\dg_z-\pi_z)&0\\
\end{array}\right)\label{defzha}
\EE
(the matrix is antisymmetric and so we only wrote down the lower diagonal).
Using Wick's theorem for the electron fields $R_{\a,i}$ and $L_{\a,i}$,
we find that the $l^{ab}$ obey an SO(5)$_1$ Kac-Moody algebra, different
from the standard SO(5) algebra by a quantum anomaly coming from the
necessity for normal ordering with respect to the vacuum:
\BE
[l^{ab}(x),l^{cd}(y)]= 
\delta(x-y)\big(\delta^{ac}l^{bd}(x)-\delta^{ad}l^{bc}(x)
-\delta^{bc}l^{ad}(x)+\delta^{bd}l^{ac}(x)\big)
+{i\over2\pi}\delta'(x-y)(
\delta^{ac}\delta^{bd}-\delta^{ad}\delta^{bc})
\label{SO5comm}\EE
A similar procedure gives the corresponding relationship for left-moving
currents.

How can the SO(5) symmetry currents be expressed in the Majorana
(vector) representation $\xi$? A vector representation of the SO(5)
generators may be easily written down:
\BE
t^{(ab)}_{ij} = i\left(\delta_i^a\delta_j^b-\delta^a_j\delta_i^b
\right)
\EE
The currents (\ref{defzha}) can then be represented as follows:
\BE
l^{ab}(x) = {1\over
2}\sum_{i,j=1..5}\xi_it^{(ab)}_{ij}\xi_j
\EE
where the 5 fermions $\xi_i$ are numbered as in Eq.~(\ref{so5vector}):
Note that the $\pi$ operators correspond to bilinears
involving one fermion $\xi$ from the spin sector and one from the charge
sector, which again fits with the physics since we know that they create
objects with both spin and charge.

In the low-energy sector of the model (\ref{TLmodel}), the spin-charge sector
can be represented by an SO(5)$_1$ WZW model, perturbed away from the perfectly
symmetric point by current-current interactions (the only interactions present
in the spin and charge sectors in our model). So Zhang's idea of using an SO(5)
$\s$-model representation with symmetry breaking interactions\cite{zhang}
is explicitly seen to be valid for this model, and the Hamiltonian for the spin
and charge sectors can be written in terms of the  SO(5) currents in the
Sugawara form, analogous  to the form proposed in Ref.~\onlinecite{zhang}:
\BEQ\label{so5def}
H_{cs} &=& H_{0s}+H_{0c}+V_c\NN\\
&=& {\pi v_F\over 4}\int dx\sum_{a<b} \left\{
(l^{ab})^2+({\bar l}^{ab})^2\right\} 
+\int dx \left\{g_2 l^{15}{\bar l}^{15}
+g_4\left((l^{15})^2+({\bar l}^{15})^2\right)\right\}
\EEQ

One notable difference between this system and that considered by Zhang is that
here we are working away from half-filling; the $\pi$ operators are defined
slightly differently, to carry momentum $(\pm 2k_F, \pi)$. The chemical
potential term in the Hamiltonian, which in 2d breaks SO(5), here (due to
perfect nesting)  simply renormalises the wavevector $k_F$; momentum is still
conserved and the algebra (\ref{SO5comm}) still closes because operators
carrying $2k_F$ only give non-zero expectation values when combined with
operators carrying $-2k_F$. If we were to work at half-filling, there would be
an additional Umklapp scattering which would lead to a Mott charge
gap.\cite{giamarchi}

Since there are $\pi$ operators in this system, one may also ask whether there
is a well-defined $\pi$-resonance as claimed in 2d.\cite{demler,zhang} If this
is so, the commutator of the Hamiltonian with the $\Pi$ operator will be
proportional to the $\Pi$ operator.\cite{zhang} It can easily be seen that this
will not be the case. Later on, in section V, we obtain bosonised forms for
these operators. Then one can see that in Fourier space, the correlator of $\pi$
operators does not have a simple pole; their effect is not to generate a single
well-defined triplet excitation but a shower of unconfined spinons and
holons.
    
\section{Bosonisation}\label{nabos}

The low-energy sector of the model (\ref{TLmodel}), i.e., the perturbed
SO(5) WZW model of Eq.~(\ref{so5def}) is exactly solvable, in the sense that
we may find the exact energy levels of low-lying states and the long-distance
correlations of various operators. We will first indicate the physical content
of the SO(5) WZW model without perturbations, and then see explicitly how the
interactions $g_2$ and $g_4$ separate spin and charge sectors, affect
correlation exponents and deform low-energy SO(5) multiplets.

In the language of conformal field theory,\cite{DMS} particularly useful when
dealing with critical theories, each WZW model contains a finite number of
primary  fields, having well-defined conformal dimensions $\Delta$ and
$\bar\Delta$. An operator $A$ with such conformal dimensions has the following
dynamical correlations:
\BE
\langle A(x,t) A(0,0)\rangle \sim
{1\over (x-vt)^{2\Delta} (x+vt)^{2\bar\Delta}}
\EE
The level-1 SO(5) WZW model has two primary fields: a five-component vector
field $\xi$ of conformal dimension $\Delta=\hf$ and a four-component spinor
field $h$ of conformal dimension $5\over16$. Under SO(5) rotations, these fields
transform respectively in the vector and spinor representations of SO(5). Of
course, the field $\xi$ is made of the five Majorana fermions
(\ref{so5vector}), whereas the field $h$ is what is left of the original
electron fields $R_\mu$, originally in a spinor representation of SO(8), after
the flavour sector has been gapped out. Freezing out the flavour part has had
the effect of decreasing the conformal dimension of the spinor field from $\hf$
(in SO(8)) to $5\over16$ (in SO(5)), thus making it more relevant. In addition
to these primary fields, the ten SO(5) currents (\ref{defzha}) also play a
crucial role in the theory and their correlations may also be exactly
calculated.

\subsection{Spin-charge separation}

When the interactions of (\ref{so5def}) are turned on, the SO(5) symmetry is
explicitly broken and the spin and charge sectors of the theory separate.
The spin sector, unaffected by the interactions, becomes a level-1 SO(3) WZW
model, which is the same as a level-2 SU(2) WZW model. The SU(2)$_2$ WZW theory
contains two primary fields: a spin triplet (or vector) $\xi^s_i$ ($i=1,2,3$)
with conformal dimension $\hf$, and a spin-$\hf$ (or spinor) field $g_\a$
($\a=\up,\dn$), of conformal dimension $3\over16$. Products of the left- and
right-moving parts of $g$ are commonly arranged in a $2\times2$ matrix:
\BE\label{WZWmatrix}
G = \pmatrix{g_\up\bar g_\up & g_\up\bar g_\dn \cr
g_\dn\bar g_\up & g_\dn\bar g_\dn \cr}
\EE

The charge sector becomes a U(1) theory, which may be described by a single
boson field $\Phi_c$. The effect of the interactions on the charge boson is
simply to change the spectrum of anomalous dimensions ($K_c\ne1$) and the
theory remains critical. Since Abelian bosonisation is fairly
standard,\cite{coleman,mandelstam,banks,stone} we shall only state a few
results. The charge boson $\Phi_c$ may be written as the sum of right and left
parts: $\Phi_c=\phi_c+\phib_c$. Defining the dual field $\theta_c$ as
$\theta_c=\phi_c-\phib_c$, the charge Hamiltonian may be written as
\BE
H_c = {v_c\over2}\int dx\left[ K_c(\d_x\theta_c)^2
+{1\over K_c}(\d_x\Phi_c)^2\right] 
\EE
where
\BEQ
K_c&=&\sqrt{\pi v_F+g_4-g_2\over \pi
v_F+g_4+g_2}\NN\\
v_c&=&\sqrt{\left(v_F+{g_4\over \pi}\right)^2
-\left({g_2\over \pi}\right)^2}
\label{ll}\EEQ
For more general lattice Hamiltonians that are Luttinger
liquids in the low energy sector, the parameters $v_c$ and $K_c$
depend in a more complicated way upon the original couplings, so in the
following analysis, one can just treat them as independent parameters
whose precise value depends upon the original model.

If $g_2=0$, there are no anomalous exponents ($K_c=1$) and the scaling
fields $e^{\pm i\sqrt{4\pi}\phi_c}$ represent right-moving fermions of
conformal dimensions $(\frac12,0)$. When $g_2$ is turned on, the original
charge current ($l^{15}=J_R$) is no longer conserved but becomes a linear
combination of currents that are still conserved (see eg.
Ref.~\onlinecite{voit}):
\BE
J_R = j_R\cosh\vartheta -j_L\sinh\vartheta \qquad (K_c=\e^{-2\vartheta})
\EE
where $j_R$ and $j_L$ have respectively conformal dimensions $(1,0)$ and
$(0,1)$. The chiral components $\phi_c$ and $\phib_c$ mix through the same
Bogoliubov transformation and the original fermion operators $e^{\pm
i\sqrt{4\pi}\phi_c}$ acquire a left conformal dimension:
\BE
\Delta = {1\over 8}\left(K_c+{1/K_c}\right)+\frac14\qquad
\bar\Delta = {1\over 8}\left(K_c+{1/K_c}\right)-\frac14
\label{anomaldim}
\EE

Although the above results are very well known, it is worth pausing over them
for a moment. They show that the interaction strengths $g_2$ and $g_4$ are not
relevant energy scales in the low energy theory. They only appear as ratios
$g_{2,4}/v_F$ in the renormalisation of the velocity $v_c$ and the anomalous
exponent $K_c$. This is a very non-perturbative result. If we recall the exact
solution of the one-dimensional Hubbard model away from  half-filling, we know
that it is a Luttinger liquid for all $U>0$ from
$0$ to $\infty$.\cite{hubb} In this range, $K_c$ varies from its
noninteracting value of $1$, to $1/2$ at $U=\infty$. Even when the
on-site repulsion is infinite, its effect in the low energy sector is
just a fairly small renormalisation of the anomalous exponent! The
theory is still critical with gapless spin and charge excitations.

Some critics of the Zhang SO(5) proposal have claimed that because of the
strong on-site repulsion in the Hubbard model, the $\pi$ operators cannot
create low energy excitations.\cite{greiter} The argument is essentially that
one is forced to put two electrons on the same site, which costs an energy of
order $U$. The reason that the criticism \cite{greiter} may be too simplistic is
first of all that it is a single-particle argument, whereas the low energy
excitations of this system are many-body collective phenomena, and secondly
that it is a short length scale argument which may have some validity in the
U.V.; but we are interested in the low energy I.R. behaviour which is quite
different in a non-Fermi liquid such as the TL liquid. Even if much of the
spectral weight is shifted to high energies there is still some at low energies
and this is what dominates the low energy theory. Given that the
two-dimensional Cuprates are examples of non-Fermi liquids, it cannot be ruled
out {\it a priori} using these arguments that even in the presence of strong
on-site repulsion, the $\pi$-operators may generate low energy excitations (at
least when one is slightly away from half-filling).

Let us then see what happens to the SO(5) currents and primary fields after
spin-charge separation. Three of the five components of $\xi$ become a
spin triplet ($\xi_s$) and the remaining two are simply 
$\cos(\sqrt{4\pi}\phi_c)$ and $\sin(\sqrt{4\pi}\phi_c)$.
Out of the 10 SO(5) currents, six -- the $\pi$
operators -- are no longer conserved currents
 and may then be expressed as products of
SU$(2)^{\rm sp.}_2$ fields with charge fields. Schematically,
\BE
\pi,\pi^\dg \sim e^{\pm i\sqrt{4\pi}\phi_c(z)}\otimes\xi_s(z)
\EE
When $K_c\ne1$, the conformal dimensions of the $\pi$ operators
are no longer $(1,0)$, but rather, from Eq.~(\ref{anomaldim}) and since the
field $\xi_s(z)$ has conformal dimensions $(\frac12,0)$,
\BE
\Delta = {1\over 8}\left(K_c+{1/K_c}\right)+\frac34\qquad
\bar\Delta = {1\over 8}\left(K_c+{1/K_c}\right)-\frac14
\label{pidimen}
\EE
Thus, the $\pi$ operators are no longer conserved currents, as expected.

As mentioned above, the spinor representation $\rep4$ of SO(5) factorizes
into a pair of SU(2) doublets of charges $\pm1$ when SO(5) is broken. The
spinor field $h$ may thus be factorized as
\BE
h(x) \sim (g_\up , g_\dn) \otimes
\pmatrix{\cos(\sqrt{\pi}\phi_c)\cr \sin(\sqrt{\pi}\phi_c)\cr}
\EE
where $g$ is the SU(2) spinor mentioned above and the boson factors have
conformal dimension $1\over4$. The decomposition described here can be
rigorously proven by checking the corresponding commutators with the currents.
We obtained it differently, by the method of affine characters (see
Ref.~\onlinecite{DMS}), which we will not explain in detail here, since the
coincidence of conformal dimensions and components is sufficiently convincing
for our purpose.

\subsection{The SO(5) order parameter}

One of the interesting operators of the SO(5) WZW model (and of its perturbed
version) is a continuum version of Zhang's five-component order parameter $n_a$
($a=1,\ldots,5$).\cite{zhang} This operator can be defined in terms of the
original electron fields. The components $n_{2,3,4}$ correspond to the
staggered magnetisation and the components $n_{1,5}$ to the d-wave
superconducting order parameter. The staggered magnetisation is defined as
\BE
{\bf n}_{{\bf Q}} = \sum_{{\bf k},\a,\b}
c_\a^\dg({\bf k}+{\bf Q})
\sb_{\a\b}c_\b({\bf k})
\EE
Picking ${\bf Q}=(2k_F,\pi)$ and using Eq.~(\ref{cont}), we find
\BE
{\bf n}_{{\bf Q}} = \sum_{k,\a,\b}
\left(R^\dg_{\a,1}\sb_{\a\b}
L_{\b,1}-R^\dg_{\a,2}\sb_{\a\b}
L_{\b,2}\right)
\label{staggeredMag}
\EE
The ${\bf Q}=(-2k_F,\pi)$ component of the magnetisation
is just the Hermitian conjugate of the above. The d-wave order parameter
${\cal D}$ is defined as
\BE
\int dx\; {\cal D}^\dg = \sum_{{\bf m},{\bf n}}g_{{\bf m},{\bf n}}
c_\up^\dg({\bf m}) c_\dn^\dg ({\bf n})
\EE
where $g_{{\bf m},{\bf n}}$ has been introduced in Eq.~(\ref{gmn}).
Taking the continuum limit, we find
\BEQ
{\cal D}^\dg &=&
\sum_{i=1,2}
\left( R^\dg_{\up,i}L^\dg_{\dn,i}+L^\dg_{\up,i}
R^\dg_{\dn,i}\right)
-R^\dg_{\up,1}L^\dg_{\dn,2}
-L^\dg_{\up,1}R^\dg_{\dn,2} 
-R^\dg_{\up,2}L^\dg_{\dn,1}
-L^\dg_{\up,2}R^\dg_{\dn,1}
\label{d-wave}
\EEQ
The combinations ${\cal D}+{\cal D}^\dg$ and $i({\cal D}-{\cal D}^\dg)$
then correspond to $n_1$ and $n_5$, respectively.

An expression of the order parameter $n_a$ in terms of the scaling 
fields $h$ or $\xi$ would be more useful, since the flavour part is not
explicitly absent from the above. Such an expression is difficult to obtain in a
systematic way from the above expressions; but one can infer what it has to be
(this result can be confirmed by Abelian bosonisation; see the appendix).
Clearly, $n_a$ should be a bilinear in $h$ or $\xi$, with equal left and right
conformal dimensions. Let us consider the following SO(5) tensor products:
\BEQ
\rep4\otimes \rep4 &=& \rep1\oplus \rep5\oplus \rep{10}\NN\\
\rep5\otimes \rep5 &=& \rep1\oplus \rep{10}\oplus \rep{14}
\EEQ
This means that a bilinear in $\xi$ (5 components) cannot transform as a
vector of SO(5), whereas a bilinear in $h$ (4 components) can. We thus seek
an order parameter of the form
\BE\label{so5spingen}
n_a = \G^a_{ij}h_i\bar h_j
\EE
where the five $4\times4$ matrices $\G^a$ must transform as a vector of SO(5)
when $h$ and $\bar h$ are acted upon by a $4\times4$ unitary representation
of SO(5). If we denote by $\ell^{ab}$ a $4\times4$ representation of the
SO(5) generators, this requirement amounts to
\BE
[\ell^{ab},\G^c] = i(\delta^{ac}\G^b - \delta^{bc}\G^a)
\EE
Experience with the Lorentz group and Dirac matrices may guide us here.
If a set of five matrices $\G^a$ obey the Clifford algebra
$\{\G^a,\G^b\} = 2\delta^{ab}$, then it is a simple matter to show that
the above commutation relations are satisfied if we define
\BE
\ell^{ab} = -{i\over 4}[\G^a,\G^b] 
\EE
Moreover, the matrices thus defined do obey the SO(5) algebra
\BE
[\ell^{ab},\ell^{cd}] = i(\delta^{ac}\ell^{bd} + \delta^{bd}\ell^{ac}
- \delta^{ad}\ell^{bc} -\delta^{bc}\ell^{ad})
\EE

Let us adopt the following representation for the Clifford algebra:
\BML
\BEQ
\G^1 &=& 1\otimes \s_3 \\
\G^2 &=& \s_1\otimes \s_2 \\
\G^3 &=& \s_2\otimes \s_2 \\
\G^4 &=& \s_3\otimes \s_2 \\
\G^5 &=& -1\otimes \s_1
\EEQ\EML
Then the charge and $S_z$ matrices take the form
\BE
\ell^{51} = \hf\pmatrix{\s_2&0\cr 0&\s_2\cr} \qquad\qquad
\ell^{23} = \hf\pmatrix{{\bf 1}&0\cr 0&-{\bf 1}\cr}
\EE
With the above generators $Q$ and $S_z$, the factorisation of the chiral field
$h$ in terms of the SU(2)$_2^{\rm sp.}$ field $g$ and of the charge boson
$\phi_c$ must be
\BEQ
h &=& (g_\up\cos(\sqrt\pi\phi_c),g_\up\sin(\sqrt\pi\phi_c),
g_\dn\cos(\sqrt\pi\phi_c),g_\dn\sin(\sqrt\pi\phi_c)  \\
\bar h &=& (\bar g_\up\cos(\sqrt\pi\phib_c),\bar g_\up\sin(\sqrt\pi\phib_c),
\bar g_\dn\cos(\sqrt\pi\phib_c),-\bar g_\dn\sin(\sqrt\pi\phib_c)
\EEQ
The explicit expression for the order parameter $n_a={\rm tr}(\G^a h\bar h)$
in terms of the spin matrix field (\ref{WZWmatrix}) and of the charge boson
is then
\BML\BEQ
n_1 =& ~~ \Tr(G) &\cos(\sqrt\pi\theta)\\ 
n_2 =& i \Tr(G\s_1) &\sin(\sqrt\pi\Phi)  \\ 
n_3 =& i \Tr(G\s_2) &\sin(\sqrt\pi\Phi)  \\ 
n_4 =& i \Tr(G\s_3) &\sin(\sqrt\pi\Phi)  \\
n_5 =& -\Tr(G)  &\sin(\sqrt\pi\theta)
\EEQ\EML
We first notice that $n_{1,5}$ form a spin singlet and are the real and
imaginary parts of the complex d-wave order parameter
$\Tr(G)\exp(-i\sqrt\pi\theta)$, whereas $n_{2,3,4}$ form a spin triplet.
We also see how the scaling dimensions of $n_{1,5}$ diverge from those of
$n_{2,3,4}$ when $K_c$ is different from unity: the fields
$\cos(\sqrt\pi\theta)$ and $\sin(\sqrt\pi\theta)$ have conformal dimensions
$\Delta=\bar\Delta =1/(8K_c)$ while $\cos(\sqrt\pi\Phi)$ and
$\sin(\sqrt\pi\Phi)$ have conformal dimensions $\Delta=\bar\Delta=K_c/8$. Thus
\BEQ
\Delta(n_1)&=&\Delta(n_5) ~=~ 3/16+1/8K_c\NN\\
\Delta(n_{2})&=&\Delta(n_3) ~=~ \Delta(n_4) ~=~ 3/16+K_c/8
\label{orderparam}
\EEQ
We may also consider the SO(5) singlet $h_i\bar h_i$, which becomes
simply $\Tr(G)\cos(\sqrt\pi\Phi)$ in this representation. This field
is conjectured to be the charge density wave (CDW) order
parameter.\cite{private}
Within this model it has the same scaling dimension as the staggered
magnetisation (or spin density wave), but in a more general model with a spin
gap the two fields could have different correlation lengths since spin singlet
and triplet states would not necessarily have the same excitation energy.

To summarize, the components of the SO(5) order parameter have power-law
correlations governed by the above conformal dimensions. When full
SO(5) symmetry is present, $\Delta(\vec{n})=5/16$; and $\vec{n}$ is
the vector primary field of the SO(5)$_1$ WZW model. When $K_c\ne1$, the SO(5)
symmetry is broken and the staggered magnetisation is less ($K_c>1$) or more
($K_c<1$) relevant than the d-wave order parameter. The behaviour of the other
possible fermion bilinears can be checked by a combination of Abelian
bosonisation and an Ising model representation of bosonic exponents -- we find
that their correlations decay exponentially as a result of the gap in the
flavour sector (for details see the appendix). Of course, since we are in one
dimension, there are no real phase transitions, just enhanced fluctuations.
Thus the fluctuations in the superspin channel are enhanced whereas other
tendencies are suppressed. If a weak inter-ladder coupling were added to form a
two- or three-dimensional system, then a mean-field treatment would lead to a
phase transition in the channel that has the highest susceptibility (i.e., the
most fluctuations), i.e. an ordered phase for the most relevant operator. Thus,
this approach predicts d-wave superconductivity for weakly coupled ladders with
attractive effective interactions ($K_c>1$). Finally, since the charge and spin
sectors of this model are described by conformal field theories, one can also
recover the finite-temperature behaviour of the correlation functions in the
standard way.\cite{DMS}

\subsection{Lowest-energy states}

In the absence of interactions and interchain coupling, the low-energy sector
of the model is especially simple: the theory is a SO(8)$_1$ WZW model. The
states fall into two representations of the Kac-Moody algebra: that of the
identity, which contains states with even charge, and that of the spinor
(electron) field $R_\mu$, which contain states of odd
charge. Remember, this is just a complicated way of representing
noninteracting fermion excited states.
 As SO(8) is
broken into SO(5)$_1\otimes$SU$(2)^{\rm fl.}_2$, these representations break
into a finite number of representations of 
SO(5)$_1\otimes$SU$(2)^{\rm fl.}_2$, as indicated in Eq.~(\ref{decomp1}).
When the flavour sector is gapped, the low lying states must all be 
flavour singlets and so many of those representations become irrelevant,
in particular all the representations of odd charge coming from the spinor
$R_{\a}$.

The only surviving Kac-Moody representation in the SO(5)$_1$ theory is that of
the identity. Such a representation contains an infinite number of energy
levels, and at each level the states fall into SO(5) multiplets. In the pure
WZW model (before spin-charge separation) the excited states may be obtained
from the vacuum by applying ladder operators associated with the SO(5)
currents. Let us explain: in a system of finite length $L$, the currents may be
Fourier expanded as follows:
\BE
l^{ab}(x) = \sum_{n} e^{2\pi i nx/L} l^{ab}_n
\EE
where the sum runs over all integers (positive and negative). From the
commutation relations of the currents, one may infer commutation relations for
the modes $l^{ab}_n$ and show that, for $n<0$, $l^{ab}_n$ is a raising
operator for the energy in the WZW model. Of course, the $l^{ab}_0$ are nothing
but the SO(5) generators and allow us to navigate within a multiplet. 

The multiplet content at each energy level may be easily obtained from the
representation theory of Kac-Moody algebras, in particular by the method of
affine characters.\cite{DMS} Schematically, in the case at hand, the multiplet
content may be expressed in terms of a spectrum-generating function $X(q)$:
\BEQ
X(q) &=& \rep1 + q\;\rep{10} + q^2(\rep{14}+\rep{10}+\rep{5}+\rep{1})
+ q^3(\rep{35}+\rep{14} + 3\cdot\rep{10} + \rep{5} + \rep1)+\cdots
\label{spectrum1}
\EEQ
where the coefficient of $q^\Delta$ indicates the multiplet content of states
with conformal dimension $\Delta$. For instance, a term like $2q^3\cdot\rep{10}$
in $X(q)$ means that the multiplet $\rep{10}$ of SO(5) occurs twice with
conformal dimension $\Delta=3$ in the right-moving sector. The full low-energy
Hilbert space is a left-right product, encapsulated in the generating function
$X(q)X(\bar q)$. For instance, the term $Nq^\Delta\rep{10}\otimes\bar
q^{\bar\Delta}\rep{\bar 5}$ stand for a left-right tensor product of
multiplets, occurring $N$ times at the energy level
$(2\pi v_F/L)(\Delta+\bar\Delta)$, with momentum $(2\pi /L)(\Delta-\bar\Delta)$
($v_F$ is the common spin and charge velocity before spin-charge separation).

The eigenvalues of $S_z$ and $Q$ and the energy of each state in the
right-moving sector may be encoded in a more general spectrum-generating
function $X(q,x,y)$:
\BE
X(q,x,y) = \sum_{\rm states} q^{\Delta}x^{2S_z}y^{Q}
\EE
The advantage of spectrum-generating functions is that tensor products translate
into ordinary products of functions, and direct sums into ordinary sums.
Anticipating spin-charge separation, it is possible to write the function
$X(q.x.y)$ as a combination of spin-charge products:
\BE
X(q,x,y) = X_{\rm sp}^{(0)}(q,x)X_{\rm c}^{(0)}(q,y) + 
X_{\rm sp}^{(2)}(q,x)X_{\rm c}^{(2)}(q,y)
\label{spectrum2}
\EE
where $X_{\rm sp}^{(j)}(q,x)$ is the spectrum-generating function
for the spin-$j$ Kac-Moody representation of SU(2)$_2$ and
$X_{\rm c}^{(0,2)}$ is the analog for the charge sector. The lowest terms of
these functions are
\BEQ
X_{\rm sp}^{(0)}(q,x) &=& 1 + q(1+x^2+x^{-2})  
+ q^2(3+2x^2+2x^{-2} + x^4+x^{-4}) + \cdots \NN\\
X_{\rm sp}^{(2)}(q,x) &=& q^{1/2}(1+x^2+x^{-2}) 
+ q^{3/2}(2+x^2+x^{-2}) 
+ q^{5/2}(4 + 3x^2+3x^{-2}+x^4 + x^{-4}) +\cdots \NN\\
X_{\rm c}^{(0)}(q,y) &=& 1 +q + q^2(2+y^4 + y^{-4}) 
+ q^3(3+y^4 + y^{-4}) +\cdots\NN\\
X_{\rm c}^{(2)}(q,y) &=& q^{1/2}(y^2 + y^{-2}) + q^{3/2}(y^2 + y^{-2})
+ q^{5/2}(2y^2 + 2y^{-2}) + \cdots 
\label{spectrum3}
\EEQ
Again, the exponent of $x$ is twice the value of $S_z$ and that of $y$ is the
charge $Q$. A term like $4x^2y^{-2}q^3$ in a (\ref{spectrum2}) would stand for
four states with $\Delta=3$, $S_z=1$ and $Q=-2$.
The charge states represented in $X_{\rm c}^{(0)}$ have charge $Q=0$
(modulo 4) and those in $X_{\rm c}^{(2)}$ have charge $Q=2$ (modulo 4).
From the above expressions and relation (\ref{spectrum2}), the full
spectrum of energies and quantum numbers may be recovered. Of course, we
must consider the left-right product $X(q,x,y)X(\bar q,x,y)$.
That the expression (\ref{spectrum2}) is a sum of products, instead of being a
simple product of spin and charge factors, means that one cannot consider the
charge and spin spectra independently: there are ``glueing conditions'' between
charge and spin states, conditions encoded in (\ref{spectrum2}).

When spin and charge separate, the energy levels shift in two ways. First,
because of different spin and charge velocities, $X_{\rm sp}^{(j)}(q,x)$
and $X_{\rm c}^{(n)}(q,y)$ become respectively
$X_{\rm sp}^{(j)}(q_s,x)$ and $X_{\rm c}^{(n)}(q_c,y)$, where 
$q_s=q^{v_s/v_F}$ and $q_c=q^{v_c/v_F}$. Second, anomalous charge exponents
change the conformal dimensions in the charged sector, whose structure deserves
a more detailed explanation: excited states in the charge sector may be
obtained either (i) by applying the creation operators associated with the
charge boson $\phi_c$ (this does not change the charge $Q$) or (ii) by applying
exponentials $\exp(iQ\sqrt{\pi}\phi_c)$ on the vacuum, where $Q$ is the
charge thus created. The generating functions in the charge sector may be
expressed as
\BE
X_{\rm c}^{(\ell)}(q,y) = X_{\rm bos.}(q)\sum_{Q=4r+\ell} q^{Q^2/8}y^Q
\EE
where $r$ runs over all integers, $\ell=0$ or 2, and $X_{\rm bos.}(q)$ is the
spectrum generating function associated with the boson creation operators only:
\BE
X_{\rm bos.}(q) = \prod_{r=1}^\infty {1\over 1-q^r} 
= 1 + q + 2q^2 + 3q^3 + 5q^4 + 7q^5 + \cdots 
\EE
When $K_c$ changes from its initial value of unity, left and right boson
creation operators mix through some Bogoliubov transformation and the conformal
dimensions associated with the exponentials of $\phi_c$ and $\phib_c$ become
\BEQ\label{dims}
\Delta(Q,\bar Q) &=& {1\over32}\left[
{1\over\sqrt{K_c}}(Q+\bar Q) + \sqrt{K_c}(Q-\bar Q) \right]^2 \NN\\
\bar\Delta(Q,\bar Q) &=& {1\over32}\left[
{1\over\sqrt{K_c}}(Q+\bar Q) - \sqrt{K_c}(Q-\bar Q) \right]^2
\EEQ
Left-right products of spectrum-generating functions in the charge sector then
become
\BE
X_{\rm c}^{(\ell)}(q_c,y) X_{\rm c}^{(\ell')}(\bar q_c,y)
= X_{\rm bos.}(q_c)X_{\rm bos.}(\bar q_c)
\sum_{\scriptstyle Q=4r+\ell \atop\scriptstyle\bar Q=4r'+\ell'}
q^{\Delta(Q,\bar Q)}\bar q^{\bar\Delta(Q,\bar Q)} y^{Q-\bar Q}
\EE
The above expression, combined with
Eqs~(\ref{spectrum2},\ref{spectrum3},\ref{dims}) allows us to extract the
energy, momentum, charge and spin of the whole low-energy sector for arbitrary
values of $v_s$, $v_c$ and $K_c$.

Let us consider, for instance, the first excited states. According to
Eq.~(\ref{spectrum1}), They fall into the multiplet $\rep{10}$ of SO(5). The
spin and charge content of such a multiplet is easily read from the
corresponding $S_z-Q$ diagram of
Fig~\ref{so5fig}. The multiplet $\rep{10}$ consists of three spin triplets
(of charge $-2$, $0$ and $2$, respectively), plus a neutral spin singlet.
After spin-charge separation and if $K_c\ne1$, the contribution of this
multiplet to the spectrum-generating function is, according to
Eqs~(\ref{spectrum2},\ref{spectrum3}),
\BE
q_c + q_s(1+x^2+x^{-2}) + q_s^{1/2}q_c^{\Delta(2,0)}\bar q_c^{\bar\Delta(2,0)}
(1+x^2+x^{-2})(y^2+y^{-2})
\EE
Thus, the energies of these states split in the following fashion:
\BEQ
E(Q=0,j=0) &=& {2\pi v_c\over L} \NN\\
E(Q=0,j=1) &=& {2\pi v_s\over L} \NN\\
E(Q=\pm2,j=1) &=& {\pi v_s\over L} 
+ {\pi v_c\over2L}\left(K_c+{1\over K_c}\right) \NN\\
\EEQ
The last of these states are in fact created by applying $\pi$ operators (see
Eq.~(\ref{pidimen})). The energy levels are proportional to the
scaling dimensions of the operators in the conformal field theory.\cite{DMS}

To conclude, the eigenstates, in particular the lowest-energy states, fall into
deformed SO(5) multiplets. The amount of deformation is exactly determined
by the renormalised charge velocity $v_c$ and anomalous charge exponent $K_c$.

\section{The Luttinger liquid Case}

As we mentioned earlier, the case of
two coupled Luttinger liquids is different to that of two Luttinger
models. Let us consider, as an example of a Luttinger liquid, the
one-chain Hubbard model at weak coupling $U\ll t$:
\BE 
H_{\rm Hub} = -t\sum_{r,\a}\left(c_{r,\a}^\dg
c_{r+1,\a}+c_{r+1,\a}^\dg c_{r,\a}\right)
+U\sum_r n_{r,\up} n_{r,\dn}
\EE
If we linearise about the right and left Fermi points as in (\ref{cont}),
and use the charge currents (\ref{density}) and the corresponding spin
currents
\BE
\jv_R=\hf\sum_{\a,\b}R^\dg_\a \sb_{\a
\b} R_{\b}
\qquad,\qquad
\jv_L= \hf\sum_{\a\b}L^\dg_\a \sb_{\a
\b}L_{\b}
\EE
we find the Hamiltonian density ($v_F\sim ta_0$):
\BE
{\cal H}_{\rm Hub} \approx  -iv_F\sum_\a\left(
R_\a^\dg\d_x R_\a - L^\dg_\a\d_x L_\a\right)
+ {U\over 4}(j_R^2+j_L^2)+{U\over 2}j_Rj_L
-{U\over 3}(\jv_R^2+\jv_L^2)-2U\jv_R\cdot\jv_L
\EE
This Hamiltonian is not equivalent to a Tomonaga
Luttinger model because the last two terms are not pure density-density
interactions. The $(\jv_R^2+\jv_L^2)$ term will only renormalise
the spin velocity $v_s$ and is not very important. However, the 
marginally irrelevant $\jv_R\cdot \jv_L$ term cannot simply be
absorbed in this way. It is this term which gives rise to logarithmic
corrections to the correlation functions in Luttinger liquids -- although
otherwise it does not drastically change their properties, hence the
utility of the Luttinger liquid concept.

As this weak coupling bosonisation suggests, the coupling constant of this term
is of the same order as the other couplings in the theory, so it cannot
necessarily be jettisoned when we consider more complicated models such as the
two-chain ladder. Let us therefore consider as a model for the generic Luttinger
liquid two-chain ladder, the Hamiltonian (\ref{TLmodel}) perturbed by marginally
irrelevant spin current interactions in each chain:
\BEQ
{\cal H}_{\rm liq}&=&{\cal H}+{\cal H}_{\rm marg}\NN\\
{\cal H}_{\rm marg}&=&-\lambda\left(\jv_{R1}\cdot\jv_{L1}
+\jv_{R2}\cdot\jv_{L2}\right)
\EEQ
where $\lambda>0$, and $\jv_{Ri},\jv_{Li}$ are
the right and left moving spin currents in chains $i=1,2$. It is
instructive to write the perturbation in terms of the Majorana (vector) fermions
$\xi_i$. We find:
\BE
{\cal H}_{\rm marg} = -{\lambda\over 2}\left(
{\bf J}_R\cdot{\bf J}_L-\sum_{i=1,2,3}(\xi^s_i\xib^s_i)
\xi^f_3\xib^f_3\right)\label{marg}
\EE
The first term is a marginally irrelevant interaction in
the total spin sector (${\bf J}_R=\jv_{R1}+\jv_{R2}$ as defined
 in (\ref{currents})). But it is the second term which is most significant.
It couples the fermions of the spin sector ($\xi^s_i$= rightmoving,
$\xib^s_i$= leftmoving, $i=1,2,3$) to one of the fermions in
the flavour sector. So the spin and flavour sectors are no longer
genuinely decoupled. 

Suppose that the flavour sector becomes gapped. Then:
\BE
\langle\xi^f_3\xib^f_3\rangle \neq  0
\EE
To first approximation, we can then replace $\xi^f_3\xib^f_3$ in (\ref{marg}) by
its expectation value, and we see that the effect of a gap in the flavour
sector is to generate a mass term for the fermions of the spin sector -- a spin
gap. This is a crude argument but it is borne out by the RG analysis of
Refs.~\onlinecite{fabrizio},\onlinecite{balf}, as well as numerical
work \cite{scalapino}
 (in the notation of Ref.~\onlinecite{fabrizio} a finite backscattering
corresponds to  $g_i^{(1)}\neq 0$) which shows the existence of strong coupling
regimes with a spin gap in a model of two Hubbard chains coupled by a small
hopping. In general it
 is hard to estimate the size of this gap. If it is large, the low
energy physics will be as described in
Refs.~\onlinecite{fabrizio},\onlinecite{dgs}. It could
be, however, that in some models (with small $\lambda$ for example) the spin
gap is very small, in which case for intermediate energy scales the behaviour
will still be described approximately by the model (\ref{TLmodel}).

\section{Conclusions}

In this paper we have studied a system of two TL models coupled by a
small interchain hopping. We have
shown that this critical (at T=0) theory can be represented much more
symmetrically than in the standard Abelian bosonisation representation
as an SO(5)$_1$ WZW
model,
 or equivalently as a system of 5 Majorana fermions, perturbed by
symmetry
 breaking interactions. We have obtained the
correlations of fermion bilinears in this theory and demonstrated that the
components of the  ``superspin'' \cite{zhang} have power law correlations, and
are enhanced,  whilst other tendencies are suppressed. Conformal field theory
allows us to obtain the exact energy levels in a finite size system and observe
how the degeneracy of the SO(5) multiplets is broken by spin-charge separation
($v_c\neq v_s$) and the presence of an anomalous exponent ($K_c\neq 1$) in the
charge sector. Except in the trivial noninteracting case, there is no
exact SO(5) symmetry.
In section VI we showed briefly how the inclusion of backscattering
results in the appearance of a spin gap.

In the light of these results, S.C. Zhang has recently shown\cite{private} that
a whole class of ladder systems with more general interactions have
Hamiltonians with microscopic SO(5) symmetry. Knowing the Hamiltonian does not
tell us about the strong coupling behaviour at low energies. But a continuous
symmetry like SO(5) cannot be spontaneously broken in one dimension and must
therefore be present in the low-energy theory as well. The latter must be
described by a SO(5) WZW model, perturbed by various primary fields, perhaps
with a critical point or line in the space of coupling constants (a conformal
field theory with Lie-group symmetry is necessarily a WZW model). This is the
simplest class of low energy theories with SO(5) symmetry in 1+1 dimensions. 

The SO(5) symmetric description of ladder models, which
are clearly related to popular models of the Cuprates, and the similarity of
the form of the theory in ladders to that proposed by S.C. Zhang
\cite{zhang} for the 2d Cuprates, is certainly encouraging and suggestive.
Nonetheless, in view of the many special features of one-dimensional theories
we are cautious about drawing more general conclusions. 


\section{Acknowledgements}

D.G.S. is grateful to I. Affleck for originally drawing his attention to the
SO(5) theory, and to S.C. Zhang, E. Demler and A.M. Tsvelik for useful 
conversations. This work is supported by NSERC (Canada) and by
FCAR (Qu\'ebec).

\section{Appendix: Abelian Bosonisation of Fermion Bilinears}

We want to know the long distance (low energy) behaviour of the correlations of
various bilinears. In Sect.~(\ref{nabos}) we demonstrated that the correlations
of the ``superspin'' order parameter could be deduced within the framework of
non-Abelian bosonisation from a careful analysis of the operators of the
conformal field theory. We can further justify this analysis and find the
behaviour of the other fermion bilinears explicitly by using Abelian
bosonisation. 

We introduce an Abelian boson $\Phi_\a^i$
for each species of fermion $i=1,2$,
$\a=\up,\dn$. Then we introduce linear charge
and spin, bonding and antibonding combinations of these fields:
\BEQ
\Phi_\a^{\pm} &=& {1\over\sqrt2}\left(\Phi_\a^1\pm
\Phi_\a^2\right)\NN\\
\Phi^i_{c,s} &=& {1\over\sqrt2}\left(
\Phi^i_\up\pm\Phi^i_\dn \right)
\EEQ
If one carefully applies Abelian bosonisation to the original
Hamiltonian (\ref{TLmodel}), taking full account of the anticommutation
factors,\cite{banks} one can identify each of these Bose fields with
two of the Majorana fermions introduced in section (4). The
$\Phi_c^+$ field is simply associated with the two charge fermions, the
$\Phi_c^-$ field represents two of the flavour fermions. $\Phi_s^+$ represents 
two of the spin fermions, and $\Phi_s^-$ comprises one flavour fermion
and one spin fermion. 

\begin{table}
\begin{tabular}{llll}
&Operator&$X$&$\Delta=\bar\Delta$\\
\hline
(a)& $R^{\dg}_{1\up}L^{\dg}_{1\dn}$&
$\Phi_s^+-\theta_c^+-\Phi_s^--\theta_c^-$&--\\
(b)& $R^{\dg}_{1\up}L^{\dg}_{2\dn}$&
$\Phi_s^+-\theta_c^+-\theta_s^-+\Phi_c^-$&
${3\over 16}+{1\over 8K_c}$\\
(c)& $R^{\dg}_{1\up}L_{1\dn}$&
$\Phi_c^+-\theta_s^++\Phi_c^--\theta_s^-$&
${3\over 16}+{K_c\over 8}$\\
(d)&$R^{\dg}_{1\up}L_{1\up}$&
$\Phi_c^++\Phi_s^++\Phi_c^-+\Phi_s^-$&
${3\over 16}+{K_c\over 8}$\\
(e)&$R^{\dg}_{1\up}L_{2\dn}$&
$\Phi_c^+-\theta_s^+-\theta_c^-+\Phi_s^-$&--\\
(f)&$R^{\dg}_{1\up}L_{2\up}$&
$\Phi_c^++\Phi_s^+-\theta_s^--\theta_c^-$&--\\
\end{tabular}\smallskip
\caption{Fermion Bilinears.\label{bilin}}
\end{table}

After bosonisation, the various fermion bilinears have the form
\BE
\hat{O} \sim  \e^{-i\sqrt{\pi}X}
\EE
where the different $X$s are given in Table \ref{bilin}.
We will briefly describe how we arrive at the long distance
behaviour of their correlations. We find straightforwardly:
\BEQ
\Delta (\e^{\pm i\sqrt{\pi}\Phi_s^+})&=&
\Delta (\e^{\pm i\sqrt{\pi}\theta_s^+})={1\over 8}\NN\\
\Delta (\e^{\pm i\sqrt{\pi}\Phi_c^+})&=&{K_c\over 8}\NN\\
\Delta (\e^{\pm i\sqrt{\pi}\theta_c^+})&=&{1\over 8K_c}
\EEQ
(The scaling dimension is $D=\Delta+\bar\Delta$ and here $\Delta
=\bar\Delta$ so $D=2\Delta$)
But the charge and spin ($-$) fields are a little more subtle.
In our model, the flavour sector acquires a gap. Since
we start with $K_c<1$ this strong coupling regime 
corresponds to the limit $K_c\rightarrow 0$ in the c$-$
sector, whence to leading approximation we can replace the
complex exponents by their expectation values:
\BEQ
\langle \e^{i\sqrt{\pi}\Phi_c^-}\rangle &\neq &0\NN\\
\langle \e^{i\sqrt{\pi}\theta_c^-}\rangle
&=&0
\EEQ
Higher order corrections will die away exponentially. Thus the 
bilinears (a), (e) and (f) in Table~\ref{bilin} die away exponentially 
and the corresponding tendencies are suppressed.

The exponents of $\Phi_s^-$ are a little more subtle since we know
that only one of the Majorana fermions to which $\Phi_s^-$ corresponds
is gapped, whilst the other remains gapless. Here, however, we can make use
of their representation in terms of the corresponding Ising order and
disorder operators.\cite{shelton,sato,i+d}
Introducing Ising order and disorder operators $\s_f,\mu_f$ 
corresponding to the Majorana flavour fermion $\xi^f_3$ and $\s_s,\mu_s$
corresponding to the spin fermion $\xi_s^3$, we can identify the following
approximate operator correspondences:
\BEQ
\cos\sqrt{\pi}\Phi_s^-&\sim& \s_f\s_s\NN\\
\sin\sqrt{\pi}\Phi_s^-&\sim& \mu_f\mu_s\NN\\
\cos\sqrt{\pi}\theta_s^-&\sim& \s_f\mu_s\NN\\
\sin\sqrt{\pi}\theta_s^-&\sim& \mu_f\s_s
\EEQ
When the flavour fermion becomes gapped, either 
$\langle\s_f\rangle=0$, $\langle\mu_f\rangle\neq 0$
or $\langle\s_f\rangle\neq 0$, $\langle\mu_f\rangle=0$,
depending upon the definitions. Thus to first approximation these
can again be replaced by their expectation values. The Ising model
corresponding to the spin fermion $\xi_s^3$ remains critical and has
scaling dimensions $\Delta=\bar\Delta(\s_s,\mu_s)=1/16$.
In this way we obtain the long distance asymptotics shown in Table~\ref{bilin}.
The only bilinears which still have power law correlations are the interchain
pairing (\ref{d-wave}), represented by (b) in Table~\ref{bilin}, and the
staggered magnetisation (\ref{staggeredMag}), represented by (c) and (d).
Thus, it is precisely the components of the unified order parameter $n_a$ that
have power law correlations, while all the other tendencies around $\pm 2k_F$
are suppressed. Note that the scaling dimensions that we find agree with those
found in Sect.~\ref{nabos} from non-Abelian bosonisation (cf.
Eq.~(\ref{orderparam})).


\end{document}